# Fundamental Molecules of Life are Pigments which Arose and Evolved to Dissipate the Solar Spectrum

Karo Michaelian[a] and Aleksandar Simeonov[b]
a) Instituto de Física, UNAM, Circuito Interior de la Investigación Científica, Cuidad Universitaria, México D.F., Mexico, C.P. 04510.
b) Independent researcher, Bigla str. 7, Skopje, The former Yugoslav Republic of Macedonia.

## Abstract
The driving force behind the origin and evolution of life has been the thermodynamic imperative of increasing the entropy production of the biosphere through increasing the global solar photon dissipation rate. In the upper atmosphere of today, oxygen and ozone derived from life processes are performing the short wavelength UVC and UVB dissipation. On Earth's surface, water and organic pigments in water facilitate the near UV and visible photon dissipation. The first organic pigments probably formed, absorbed, and dissipated at those photochemically active wavelengths in the UVC that could have reached Earth's surface during the Archean. Proliferation of these pigments can be understood as an autocatalytic photochemical process obeying non-equilibrium thermodynamic directives related to increasing solar photon dissipation rate. Under these directives, organic pigments would have evolved over time to increase the global photon dissipation rate by; 1) increasing the ratio of their effective photon cross sections to their physical size, 2) decreasing their electronic excited state life times, 3) quenching radiative de-excitation channels (e.g. fluorescence), 4) covering ever more completely the solar spectrum, and 5) dispersing into an ever greater surface area of Earth. From knowledge of the evolution of the spectrum of G-type stars, and considering the most probable history of the transparency of Earth's atmosphere, we construct the most probable surface solar spectrum as a function of time and compare this with the history of molecular absorption maxima obtained from the available data in the literature. This comparison supports the thermodynamic dissipation theory for the origin of life, constrains models for Earth's early atmosphere, and sheds some new light on the origin of photosynthesis.

## 1. Introduction
Like all irreversible processes, life must have arisen to dissipate a generalized thermodynamic potential. By far the most important potential that life dissipates today is the solar photon potential. Living systems reduce the albedo of Earth and dissipate, through many coupled irreversible processes, shortwave incoming radiation into long-wave radiation which is eventually returned to space, ensuring an approximate energy balance in the biosphere. We have suggested that the optimization of this entropy production under the solar photon potential provides the motive force behind the origin and evolution of life (Michaelian, 2005; 2009; 2011; 2012a; 2012b; 2013).

Most of the earliest organic molecules, those common to all three domains of life (Bacteria, Archaea and Eukaryote), are pigments which absorb light in the far ultraviolet (UVC) and when in an aqueous environment dissipate this light efficiently into heat. Over the history of life on Earth, organic pigments have evolved to absorb in the range where water does not, approximately 220 to 700 nm. Stomp et al. (2007) have, in fact, demonstrated just how neatly organic pigments are filling photon niches left by water.

From this thermodynamic perspective, the origin of life began with the photochemical formation of organic pigments to dissipate the solar photon potential arriving at Earth's surface where water does not absorb. Specifically, we have postulated (Michaelian, 2009, 2011) that life began dissipating UVC photons within the range of 240 nm to 280 nm, where a window existed in the primitive Earth atmosphere (Sagan 1973) and where the primary molecules, those common to all three domains of life (RNA and DNA, the aromatic amino acids, and enzymatic cofactors) absorb and dissipate strongly when in water.

This "thermodynamic dissipation theory for the origin of life" suggests that the evolutionary history of life on Earth is, and always was, driven by increases in the global photon dissipation rate of the biosphere. This is obtained through optimizing the pigment photon dissipation at the enthropically most important photon wavelengths (short wavelengths) arriving at Earth's surface by evolving; 1) increases in the photon absorption cross section with respect to pigment physical size, 2) decreases in the electronic excited state life times of the pigments, 3) quenching of the radiative de-excitation channels (e.g. fluorescence), 4) greater coverage of the solar spectrum, and 5) pigment proliferation and dispersion over an ever greater surface area of Earth by evolving mobile organisms that spread essential nutrients and seeds into inhospitable environments, including mid-ocean and extreme land environments (Michaelian, 2009; 2011).

The earliest organic pigments on Earth's surface were probably formed directly via photochemical reactions on prebiotic molecules such as $H_2$, $N_2$, $CO_2$, $CH_4$, HCN, $H_2O$ and common polyaromatic hydrocarbons (Oró and Kimball, 1961). Contemporary organic pigments have to be formed through more indirect metabolic routes, since high energy photochemically active wavelengths are no longer available at Earth's surface, but are still ultimately based on photochemical reactions. When in water these pigments dissipate the solar photon potential into heat and their formation is therefore an autocatalytic dissipative process driven by entropy production. This non-linear, non-equilibrium process results in concentrations of the pigments orders of magnitude beyond their expected equilibrium values (this has been well studied for chemical reactions, Prigogine (1967), and we have extended it to the photochemical reactions, Michaelian (2013)) and thus explains the proliferation of organic pigments over Earth's entire surface (Michaelian, 2011; 2013). This proliferation, in essence, is the hallmark of biological evolution.

Driving the proliferation and evolution of life is thus the thermodynamic imperative of increasing the global entropy production of the biosphere. Since the bulk of the entropy production on Earth's surface consists mainly in the dissipation of solar photons by organic pigments in water, the history of organic pigments should correlate with the evolution of the surface solar photon spectrum. Based on this conjecture, in this article we first consider the most probable evolution of the solar spectrum at Earth's surface and compare this with a pigment history (a pigment tree of life) reconstructed from the available data in the literature. This comparison lends strong support to the thermodynamic dissipation theory for the origin of life (Michaelian, 2009, 2011), constrains models of Earth's early atmosphere, and sheds new light on the origin of photosynthesis.

**2. Evolution of earth's surface photon spectrum**

From the best available knowledge of the evolution of solar type stars and of the evolution of Earth's atmosphere we reconstruct here the most probable solar photon spectrum reaching Earth's surface as a function of time since the beginnings of life. In section 4, this reconstruction is compared with the available data concerning the history of pigment maximal absorption presented in section 3.

## 2.1 Evolution of the solar spectrum

Through studying nearby G0-V main sequence stars similar to our sun, which have known rotational periods and well-determined physical properties, including temperatures, luminosities, metal abundances and ages, Dorren and Guinan (1994) have been able to reconstruct the most probable evolution of our sun's characteristics over time, in particular, the evolution of its spectral emission. This "Sun in Time" project has been carried out using various satellite mounted telescopes including ROSAT, Chandra, Hubble, and EUVE and now has representative data for our sun's main sequence lifetime from 130 Ma to 8.5 Ga.

Over the lifetime of G0-V main sequence stars, emitted wavelengths of shorter than 150 nm originate predominantly in the chromosphere and corona (stellar atmosphere) in solar flares resulting from magnetic disturbances. The high rotation rate of a young star gives it an initially large magnetic field, before significant magnetic breaking sets in, and thus intense and more frequent solar flares leading to large fluxes of these very high energy photons. In figure 1 the short wavelength flux intensities as a function of the age of a G0-V type star are given. From the figure it can be seen that our sun at 500 Ma (at the probable beginnings of life on Earth at 3.85 Ga), would have been 5 to 80 times (depending on wavelength) more intense at these very short wavelengths.

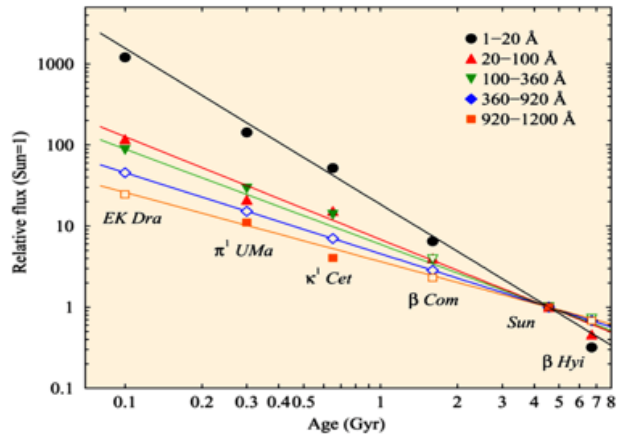

Fig. 1: Solar-normalized fluxes (with respect to those of today) vs. stellar age for different wavelength bands for solar-type stars. Taken from Ribas et al. (2005).

Wavelengths greater than 150 nm are known to originate on the photosphere and are emitted essentially in a black-body spectrum (apart from a few strong stellar atmospheric absorption lines) at the star's effective surface temperature (Cnossen et al., 2007). The effective surface temperature of a star is related to its visible luminosity $L_S$ and radius $r$ by the Stefan-Boltzmann law,

$$T_{eff} = \left(\frac{L_S}{4\pi\sigma r^2}\right)^{1/4},$$

where $\sigma$ is the Stefan-Boltzmann constant.

The luminosity of a star is an increasing function of its age because hydrogen fusion begins predominantly in the core and gradually proceeds outwards as helium "ash" settles into the core (Karam, 2003). Our sun, at the time of the origin of life, is thus expected to have been about 30% less luminous than today. The radius of a star is also a monotonically increasing function of its age (Bahcall et al., 2001). The net result for our Sun is that its effective surface temperature has been increasing steadily and at the origin of life was approximately 1.6% less than today, while the integrated UV flux (176-300nm) was most likely about 15% less than today (Karam, 2003).

**2.2 Evolution of the transparency of Earth's atmosphere**
Far more uncertain, but much more important than the incident solar spectrum in determining the solar photon flux at Earth's surface in prior times, is the wavelength dependent extinction in the atmosphere due to both absorption and scattering. Although the exact history of the primitive atmosphere and its evolution to its present state is still under active debate, there are, fortunately, well established geochemical data from the abiotic and biotic fossil record that constrain model scenarios.

Earth's second atmosphere of the Archean (the first atmosphere was probably lost to space during the heavy asteroid bombardment of between 4.4 and 4.2 Ga with a late spike in bombardment at between 3.9 and 3.8 Ga (post late-lunar bombardment; Gomes et al., 2005) is thought to have been composed mainly of $N_2$, $CO_2$, $H_2O$, some $CH_4$, $NH_3$, with the ratio $CH_4/CO_2 \ll 1$ (Kasting et al., 1983; Kharecha et al., 2005), and perhaps up to 0.1 bar of $H_2$ (Tian et al., 2005), but with very little oxygen and therefore essentially no ozone. For commonly assumed Archean atmospheric pressures of around 1 bar, these gases are all nearly perfectly transparent above approximately 220 nm (Sagan, 1973). However, considering the probable volcanic expulsion of hydrogen sulfide (the thermodynamically stable sulfur-containing gas under reducing conditions; Miller, 1957; Sagan and Miller, 1960) from many active vents and the likely formation rates of aldehydes (formaldehyde and acetaldehyde) through UV photochemical reactions on hydrogen sulfide, Sagan (1973) calculated a 240 to 290 nm window of transparency in the UV for Earth's atmosphere at the beginning of life. The two above cited aldehydes would have caused strong extinction from 290 nm to approximately 320 nm (Sagan, 1973).

The oxygen isotope content in zicron crystals betrays the existence of liquid water on Earth's surface since well before the beginning of life (Mojzsis et al., 2001). However, as stated above, the standard solar model and observational data on stars similar to our sun suggest that the wavelength integrated luminosity of our sun was as much as 30% less at the beginnings of life (Sagan and Chyba, 1997) which is inconsistent with the presence of liquid water unless one accepts a strong greenhouse atmosphere or a very different evolutionary model for our sun; such as an early high mass loss rate (Guzik et al., 1987) or a pulsar centered sun (Michaelian and Manuel, 2011). The most probable greenhouse gases are carbon dioxide, water vapor, methane, and ammonia. Low and Tice (2004) have presented geologic data suggesting that the Earth was kept warm by $CO_2$ and $CH_4$ at ratios of $CH_4/CO_2 \ll 1$, maintaining surface temperatures around 80°C at 3.8 Ga (Knauth, 1992; Knauth and Lowe, 2003) and falling to $70 \pm 15$ °C at 3.5–3.2 Ga (Low and Tice, 2004). Ammonia is very susceptible to UV destruction and therefore it is not considered to have been a major component of Earth's atmosphere during the Achaean (Sagan, 1973).

Using a model based on the present spectral emission of the G5 star $\kappa^1$ Cet (HD 20630) of an estimated age of 0.5 Ga, the age of the Sun at the beginning of life (3.95-3.85 Ga), and the Beer-Lambert law for atmospheric absorption but without considering multiple scattering, Cnossen et al. (2007) have shown that wavelengths below about 200 nm would have been strongly extinguished by $N_2$, $H_2O$ and $CO_2$ in the atmosphere. In their estimation, at no time in Earth's history would its surface have been subjected to radiation shorter than 200 nm. There is, however, some contradictory evidence in mass-independent sulfur isotopic signatures, suggesting that wavelengths as short as 190 nm may have penetrated to low altitudes in the Earth's early atmosphere, at least until about 2.45 Ga (Farquhar et al., 2001).

Cnossen et al.'s (2007) calculations for the photon flux intensity at 260 nm, where RNA and DNA absorb strongly, is approximately $1 \times 10^{-2}$ W m$^{-2}$ nm$^{-1}$ (or $1.4 \times 10^{12}$ photons s$^{-1}$ cm$^{-2}$ nm$^{-1}$) which, although seemingly small, is roughly $10^{31}$ times greater than what it is today due to the absorption at this wavelength by ozone and oxygen in today's atmosphere (Cnossen et al., 2007). Considering only absorption based on the predicted amount of oxygen in the atmosphere, and the expected cross section for photochemical production of ozone, Karam (2003) has predicted a somewhat higher photon flux at 260 nm of approximately $1 \times 10^{13}$ photons s$^{-1}$ cm$^{-2}$ nm$^{-1}$. Sagan (1973) has calculated an integrated flux of light reaching Earth's surface of wavelength $\leq$ 290 nm of approximately 3.3 W m$^{-2}$ and has estimated that such a flux would be lethal to today's organisms in under 0.3 s.

Although $CO_2$, $H_2O$ and the other primitive gases would have absorbed almost all of the very high energy incident photons (<200 nm) coming from an early sun (see figure 1) of age <0.5 Ga, this light still has some relevance to the spectrum at Earth's surface since some photons would have Compton scattered into energies within the atmospheric window of transparency of between 240 and 290 nm and be available at the surface of Earth for absorption by the early organic pigments. Employing calculations by Smith et al. (2004), Cnossen et al. (2007) estimate that the contribution of Compton scattered X-ray photons into an atmospheric window of 200-300 nm would have provided an additional ~$10^{-6}$-$10^{-5}$ Wm$^{-2}$nm$^{-1}$, with the scattering of still higher energy gamma rays into the window increasing this only slightly.

Low and Tice (2004) have suggested a gradual depletion of atmospheric $CO_2$ through the formation with silicates of carbonates starting around 3.2–3.0 Ga through weathering of the newly forming continental crust, including the Kaapvaal and Pilbara cratons. By 2.9–2.7 Ga, $CH_4$/$CO_2$ ratios may have become ~1 thereby stimulating the formation of an organic haze that would have given rise to a large visible albedo and reducing surface temperatures to below 60°C at 2.9 Ga, perhaps allowing oxygenic photosynthetic organisms to thrive and increasing the amount of oxygen and ozone in the atmosphere (Low and Tice, 2004). Pavlov et al. (2000) have shown that 1,000 ppmv each of $CH_4$ and $CO_2$ (Kharecha et al., 2005) would counteract the faint young sun sufficiently to keep temperatures above freezing at this time. However, not all of Earth may have remained above freezing since glacial tillites have been identified in the ~ 2.9 Ga Pongola and Witwatersrand Supergroups of South Africa (Young et al., 1998; Crowell, 1999). Eventual erosion of the continents and tectonic recycling of $CO_2$ would have allowed the $CH_4$/$CO_2$ ratio to reduce again, bringing back a warm greenhouse atmosphere to the late Archean.

Using as a model the results obtained from the Cassini/Huygens mission for Titan, Trainer et al. (2006) have investigated the probable formation of organic haze on an early Earth through photolysing of $CH_4$ with the solar Lyman-$\alpha$ line at 121.6 nm in a $N_2$ and $CO_2$ atmosphere. In laboratory experiments designed to simulate Earth's atmosphere at the origin of life, in which the $CH_4$ mixing ratio was held at 0.1%, and the $CO_2$ mixing ratio was varied from 0 to 0.5% (suggested to include most reasonable estimates for the early Earth; Pavlov et al., 2000), they found principally molecules of mass over charge ratio ($m/z$) around 39, 41, 43, and 55 indicative of alkane and alkene fragments. The amount of aromatics of 77 and 91 amu decreased with increasing $CO_2$. The C/O ratio rather than the absolute concentrations of $CH_4$ and $CO_2$ was shown to be the factor most correlated with the chemical composition of the products. Aerosol production was seen to be maximum at C/O ratios close to 1, which according to Low and Tice (2004) would have occurred at approximately 2.9 Ga. Approximately spherical particles were found in the experiments with average diameters of about 50 nm. Particles of this size, at the estimated photochemical production rates, would have produced an optically thick layer in the UV but a rather thin layer in the visible (Trainer et al., 2006). However, as observed on Titan, and in laboratory experiments employing an electrical discharge source, these particles readily form fractal aggregates of size > 100 nm, consistent with observations of the atmosphere of Titan, thereby increasing significantly the visible attenuation with respect to the UV (Trainer et al., 2006).

Given the intensity of the Lyman-$\alpha$ line (121.6 nm) from hydrogen in the sun at Earth's upper atmosphere and the probable concentrations of $CH_4$ at these altitudes, Trainer et al. (2006) estimate an aerosol production rate on early Earth of between $1 \times 10^{13}$ and $1 \times 10^{15}$ g year$^{-1}$, which is comparable or greater than the estimated delivery of pre-biotic organics from hydrothermal vents and comet and meteorite impacts combined.

By studying the sulfur isotope record, Domagal-Goldman et al. (2008) suggest that a thick organic haze, which blocked UV light in the 170-220 nm range from the photolysing of $SO_2$ in the lower atmosphere, arose at 3.2 Ga and persisted until 2.7 Ga. Based on these isotope ratios, they suggest that Earth's atmosphere went from a hazeless to thick haze between 3.2 and 2.7 Ga, and then again to a thin haze after 2.7 Ga. The appearance of the haze may be associated with the appearance of methanogens (and anoxygenic photosynthesizers) around 3.2 Ga, which led to a buildup of $CH_4$, while continent erosion led to a decline in $CO_2$, and as the ratio of $CH_4/CO_2$ became close to one, the organic haze became thicker and spread over the upper atmosphere.

A new study (Crowe et al., 2013) suggests that there were appreciable levels of atmospheric oxygen ($3 \times 10^{-4}$ times present levels) about 3 billion years ago, more than 600 million years before the Great Oxidation Event and some 300–400 million years earlier than previous indications for Earth surface oxygenation. The researchers suggest that the observed levels are about 100 000 times higher than what can be explained by regular abiotic chemical reactions in Earth's atmosphere; therefore, the source of this oxygen was almost certainly biological.

There is evidence of oxygenic photosynthesis by at least ~ 2.78 Ga in the presence of 2-$\alpha$ methyl hopanes from $O_2$-producing cyanobacteria (Brocks et al., 1999) and sterols from $O_2$-requiring eukaryotes (Summons et al., 1999) in sediments of this age (Brocks et al., 2003). The buildup of oxygen consumed the $CH_4$ in the atmosphere leading to a reduction in the organic haze. The oxygenation of Earth's atmosphere may have begun in earnest at about 2.9 Ga but accelerated at about 2.45 Ga and would have removed most of the $CH_4$ greenhouse gas from the atmosphere by about 2.2-2.0 Ga (Rye and Holland, 1998).

Other lines of geochemical evidence suggest that the major oxygenation event occurred in the atmosphere at about 2.2 Ga, with atmospheric $O_2$ levels rising sharply from < 1% of present atmospheric levels to about 0.15 of present atmospheric levels, during a narrow window of time from 2.2 to 2.1 Ga (Nisbet and Sleep, 2001; Wiechert, 2002). The sulfur isotope record and carbon deposition rates suggest that a second sharp rise in atmospheric $O_2$ approaching present levels occurred around ~ 0.6 Ga (Canfield and Teske, 1996). But during the time period from 2.2 to ~ 0.6 Ga, where atmospheric oxygen levels were about 0.15 of present levels, deep ocean water was, according to recent findings, still anoxic and furthermore highly sulfidic (Anbar and Knoll, 2002).

The spawning of wildfires requires an atmospheric oxygen content of at least 13% and the first evidence of charcoal deposits comes from the Silurian at 420 Ma (Scott and Glasspool, 2006). Recent results from the analysis of plant material trapped in amber suggest that oxygen levels did not rise to present day levels of 21% by mass until very recently, remaining at levels of between 10 and 15 % at least from 250 Ma to 30 Ma (Tappert et al. 2013).

Through an analysis of the imprint of "fossil raindrops" from 2.7 Ga discovered in Ventersdorp in the North West Province of South Africa, Som et al. (2012) have concluded that atmospheric pressures in the Archean was probably similar to today's and certainly no more than twice as large as today.

The amount of water in the present day hydrologic cycle, and thus in today's atmosphere, has been predicted to rise by about 3.2% for every 1°K increase in surface temperature due to greenhouse warming (Kleidon and Renner, 2013). Another determination can be made from the saturation pressure of water which increases about 6.5% per degree. With temperatures in the Archean at least 50 °C above those of today, a conservative estimate for the amount of water vapor in the atmosphere would be at least twice as large as today.

### 3. The organic pigment tree of life

The evolutionary history of organic pigments is a subject barely considered hitherto but crucial for framing life and evolution within the proposed thermodynamic context. From this thermodynamic perspective, one would expect the history of pigment appearance and evolution to be correlated with the evolution of the solar spectrum at Earth's surface.

It is probable that the evolution of organic pigments would have kept pace with the evolving transparency of Earth's atmosphere and that there should be a continuity over time of solar photon absorption and dissipation, reflecting the Earth's surface solar spectrum, from the earliest molecules of life (nucleotides, amino acids, cofactors) to present-day phototropic organisms with their extensive array of pigments. It is proposed here that an important hallmark of the evolution of life on Earth is the proliferation of organic pigments over Earth's surface and their structuring in response to changes of the entropically most intense part of the surface solar spectrum. This continuous history of pigment proliferation and re-structuring can be resolved into stages, and we chose six basic stages which correlate with particular geological eons supporting particular atmospheric transparencies.

1.  World of earliest pigments (Hadean eon, ~ 4.6 – 4.0 Ga)

The presence of both liquid and gaseous water phases 4.3 billion to 4.4 billion years ago (Mojzsis et al., 2001) implies the existence of a primitive water cycle with the probable involvement of organic pigments (Michaelian, 2009). It is generally believed that various pre-biotic inorganic and organic substances such as $H_2$, $N_2$, HCN, $CO_2$, $CH_4$ and polyaromatic hydrocarbons were abundant in this early aquatic environment and that UV light and lightning acting on these molecules could have led to more complex organics (Miller S., 1953; Cleaves et al., 2008). Whether the origin of the prebiotic organic material was predominantly terrestrial or extraterrestrial is irrelevant to the main premise of this paper. Complex organic molecules are, in fact, ubiquitous throughout the cosmos and are formed basically by the action of UV light on simpler prebiotic organic and inorganic carbon compounds (Callahan et al., 2011; Kwok, 2009; Gerakines et al., 2004).

Many species of organic molecules (especially those with conjugated systems – Zollinger, 2003) when dissolved in water are very efficient absorbers and dissipaters of UV and visible photons and should therefore be considered as organic pigments during the Hadean and Archean eons. Fast internal conversion in polyatomic molecules is a result of the coupling of the electronic modes to the vibrational modes, in particular through those attributed to the OH- and CH- groups (Vekshin, 2002). The vibrational energy is then dissipated and channeled via intermolecular vibrational energy transfer to nearby hydrogen bonds between water molecules fomenting evaporation.

These properties would have given them excellent entropy producing capacity. It is, therefore, probably, not a coincidence that the fundamental molecules of life (nucleic acids, aromatic amino acids, enzymatic cofactors) are actually organic pigments in the UVC range. These molecules, that are at the foundations of life today, could have floated on the surface of the primordial ocean in monomeric or single strand polymeric form, absorbing and dissipating UV and visible solar photons, transferring this energy to surrounding water molecules and enhancing in this manner the early water cycle and Earth's global entropy production in its solar environment (Michaelian, 2012a).

In support of this conjecture is the fact that the five naturally occurring nucleic bases (adenine, guanine, thymine, cytosine, uracil) are UV-pigments which absorb in the 240-290 nm UV region, the short wavelength part of the solar spectrum which could have filtered through the Earth's early Hadean and Archean atmosphere (Sagan, 1973), dissipating this radiation into heat with subpicosecond singlet state lifetimes (Pecourt et al., 2000; Middleton et al., 2009). Another interesting and corroborating fact is that the non-natural tautomers of the nucleic acid bases have much longer decay times (Serrano-Andrés and Merchán, 2009; Gustavsson et al., 2006; Crespo-Hernández et al., 2004) implying much lower entropy producing efficacy and photochemical stability. Furthermore, it has been demonstrated that the guanine-cytosine Watson-Crick base pairs exhibit superior photon-dissipation characteristics compared to other possible base pairing between these nucleobases (Abo-Riziq et al., 2005). Woutersen and Cristalli (2004) have found that vibrational relaxation of the NH-stretching mode occurs much faster in the adenine-uracil Watson-Crick base pairs compared to isolated monomeric uracil and that the hydrogen bonding between these bases is responsible for the extremely fast vibrational relaxation in the base-pair system. Therefore, the coupling of nucleobase photon dissipation to the water cycle and the resulting increase in entropy production can be seen as the thermodynamic driving force for their polymerization into RNA/DNA and the ensuing replication of these polymers through an ultraviolet and temperature assisted RNA/DNA reproduction (UVTAR; Michaelian, 2009; 2011).

Other complex issues, like the homochirality of life (right-handedness of the natural RNA and DNA enantiomers and the left-handedness of amino acid enantiomers) and an explanation for the beginnings of information storage within DNA (related to the chemical affinity of aromatic amino acids to nucleobase sequences) can also be resolved within the scope of this thermodynamic view (Michaelian, 2011).

In addition to nucleic acids, various other organic UVC-absorbing pigments could have proliferated on the surface of the Hadean ocean driven to high concentration by their photon dissipation capabilities (Michaelian, 2013). UVC light must have acted as a selective force (Sagan, 1973; Mulkidjanian et al., 2003; Serrano-Andrés and Merchán, 2009), not only for stability, but for photon dissipation characteristics and this was probably the onset of an evolution through a thermodynamic natural selection. Among the contemporary, more universal biological molecules, aromatic amino acids (phenylalanine, tyrosine, tryptophan, histidine), cystine (Pace et al., 1995; Edelhoch, 1967); pteridine (pterin and flavin), pyridine, quinone, porphyrin and corrin cofactors; as well as different isoprenoids, all absorb in the UVC, and some of them in the visible as well (see table 1).

| Common pigments to all 3 domains of life | Absorbance maximum (nm) | Molar extinction coefficient (cm$^{-1}$M$^{-1}$) |
|---|---|---|
| Adenine† | 261.0 | 13400 |
| Guanine† | 243.0 | 10700 |
| Thymine† | 263.8 | 7900 |
| Cytosine† | 266.50 | 6100 |
| Uracil† | 258.3 | 8200 |
| Phenylalanine‡ | 257.5 | 195 |
| Tyrosine‡ | 274.2 | 1405 |
| Tryptophan‡ | 278.0 | 5579 |
| Histidine‡ | 211.0 | 5700 |
| Thiamine* | 235, 267 | 11300, 8300 |
| Riboflavin¥ | 268, 373 | 31620, 10000 |
| Folic acid° | 256, 283, 368 | 26900, 25100, 9120 |
| Nicotinamide• | 262 | 2780 |
| Pyridoxine◊ | 253, 325 | 3700, 7100 |
| Ubiquinone-10ǂ | 275 | 14240 |
| Phytomenadione# | 248 | 18900 |
| Hydroxocobalamin¶ | 361 | 27500 |
| Heme A∞ | 430, 532, 584 | 118000, 8200, 27000 |

**Table 1:** Molar extinction coefficients at maximum absorption of the nucleobases, aromatic amino acids and enzymatic cofactors common to all three domains of life.
† Fasman, 1975, ‡ Fasman, 1976, * Sigma-Aldrich, ¥ Sigma, ° Sigma, • McLaren et al., 1973, ◊ Glick, 1964, ǂ Podda et al., 1996, # Nollet et al., 2012; Suttie, 2007, ¶ Hill et al., 1964, ∞ Caughey et al., 1975.

It is conceivable that the pre-cofactor function of these molecules was absorption and dissipation of the available UV and visible solar photons. There is consistency in that they all absorb from 220 nm to 290 nm and beyond 320 nm but not between 290 and 320 nm as would be expected if the atmospheric aldehydes were absorbing there (Sagan, 1973). Apart from their UVC absorbing characteristics, pterin and flavin coenzymes are known to be photochemically active chromophores in a number of actual photoenzymes and sensory photoreceptor proteins (Kritsky et al., 1997; 2010), suggesting a continuity of function since their first appearance near the origin of life. Examples are pterin- and flavin-based photoreceptor proteins called cryptochromes and phototropins in plants which mediate plant phototropism (Brautigam et al., 2004) and the pigment cyanopterin which functions as UV/blue photoreceptor in cyanobacterial phototaxis (Moon et al., 2010). The cofactor component of these photoreceptor proteins is actually responsible for the absorption of incident visible and UV photons and they have even been proposed as ancient pre-chlorophyll photosynthetic pigments (Kritsky et al., 2013; 2013). The amino acid tryptophan has been found to play a similar role of sensing UV-B light in the UVR8 plant photoreceptor (Christie et al., 2012).

Several abiotic syntheses routes for the nucleobases and cofactors have been devised using UV light. Powner et al. (2009) have shown the feasibility of ribonucleotide production under plausible prebiotic condtions (bypassing the difficult production of ribose and free pyrimidine nucleobases separately) by employing UV light at 254 nm and a heating and cooling cycle to enhance ribonucleotide synthesis over other less endergonic products. Bernstein et al. (1999) have obtained quinones from polycyclic aromatic hydrocarbons (PAHs) in water ice which were exposed to ultraviolet radiation under astrophysical conditions. Nicotinamide (Dowler et al., 1970; Cleaves and Miller, 2001), pyridoxal (Austin and Waddell, 1999) and flavin-like compounds (Heinz et al., 1979; Heinz and Ried, 1981; Heinz and Ried, 1984) have also been obtained in experiments under prebiotic conditions.

Pigment-catalyzed evaporation may have led to increased concentration of organics and ions in the sea surface layer thus promoting molecular interactions. In this manner an early association through chemical affinity between the replicating nucleic acids and other pigment molecules is plausible. Yarus et al. (2009) have provided experimental evidence for a stereochemical era during the evolution of the genetic code, relying on chemical interactions between amino acids and the tertiary structures of RNA binding sites. According to their results a majority (approximately 75%) of modern amino acids entered the code in this stereochemical era; however, a minority (approximately 21%) of modern codons and anticodons were assigned via RNA binding sites that existed during this stereochemical era. Interestingly, the binding site for the aromatic amino acid tryptophan is among the simplest of the amino acid binding sites known, as well as selective among hydrophobic side chains, and there is a recurring CCA sequence (a tryptophan anticodon triplet) which apparently forms one side of the binding site (Majerfeld and Yarus, 2005). Moreover, Polyansky and Zagrovic (2013) have provided strong evidence for the stereochemical foundation of the genetic code and suggest that mRNAs and cognate proteins may in general be directly complementary to each other and associate, especially if unstructured. Their results show that of all 20 proteinogenic amino acids, tryptophan has the highest binding affinity for purines, and histidine has the highest binding affinity for pyrimidines, followed by phenylalanine and tyrosine.

Toulmé et al. (1974) tested the binding of tryptophan-containing small peptides to heat-denatured and UV-irradiated DNA, which resulted in direct stacking interaction between the indole ring of tryptophan and single-stranded regions of DNA, that lead to tryptophan fluorescence quenching, and also a preferential binding of the peptide to thymine dimers which photosensitized the splitting of the dimer, possibly acting as a primordial photolyase enzyme. Quenching of molecules by other molecule is accomplished through the population of the vibrational levels of the quencher molecule, in much the same was as internal conversion on single molecules (Vekshin, 2002).

Mayer et al. (1979) have tested oligopeptides containing tyrosyl, lysyl, and alanyl residues which bind to polynucleotides and found that tyrosyl fluorescence of the peptides is quenched in their complexes with both single-stranded and double-stranded nucleic acids. An energy transfer mechanism from tyrosine to nucleic acid bases was proposed to account for fluorescence quenching in oligopeptide complexes with double-stranded DNAs. They further theorize that due to the specificity of its stacking interaction for single-stranded nucleic acid structures, tyrosine might be involved through such interactions in the selective recognition of single strands by proteins.

These data give clues as to the origin of the genetic code. It's plausible that the initial association of nucleic acids was with small peptides containing aromatic amino acids which bound preferentially to specific DNA/RNA sites, thus forming symbiotic-like systems in which nucleobases provided fluorescence quenching through internal conversion of the excited aromatic amino acid residues, and the peptide provided a larger UVC photon absorption cross section and primitive enzymic functions, like the splitting of thymine dimers. Apart from the increased dissipation of the coupled system, the larger cross section afforded by aromatic amino acids and cofactor molecules would produce more local heating, favoring denaturation of DNA/RNA in an ever colder sea, thus fomenting replication and so maintaining and even incrementing entropy production through photon dissipation (Michaelian, 2009; 2011).

2. World of primitive pigment complexes (early Archean eon, ~ 4.0 - 3.8 Ga)
The cooling of the ocean at the beginning of the Archean probably became the second major selective force after photon dissipation in organic pigment evolution, which induced further association of the aromatic amino acids and other pigment molecules with DNA or RNA and growth in complexity of these structures. In order for denaturing, replication and resulting proliferation of the UV-absorbing DNA/RNA polymers to persist in ever colder waters, their auxiliary set of light-harvesting antenna pigments (small peptides and cofactor molecules) had to enlarge and grow in complexity by acquiring primitive enzymatic functions. Michaelian (2011) argues how the thermodynamic imperative of maintaining and increasing the entropy production in ever colder seas led to the origin of the information content and the associated need for reproductive fidelity of RNA/DNA. Toulmé et al. (1974) have shown how tryptophan-containing peptides might have played an enzymic role in the maintenance of this fidelity by acting as primitive photolyase enzymes. Goldfarb et al. (1951) have estimated that each peptide bond contributes an average of about 2500 to 2800 to the molar absorption coefficient of proteins at 250 nm, and this was probably the thermodynamic reason for the pre-biotic synthesis of polypeptides and their association with RNA and DNA under high UVC light conditions of the early Archean.

By this stage, one can imagine the existence of primitive virus-like vesicles, made of nucleic acid interior (core) and an envelope (shell) made of enzymically-active, but still photon dissipating, proteins and other smaller pigment molecules, with simple metabolic reactions between them and the surrounding dissolved ions and molecules. Biochemist Sidney Walter Fox and co-workers (Fox and Kaoru, 1958) have synthesized protein-like chains dubbed 'proteinoids' from a mixture of 18 common amino acids at 70°C in the presence of phosphoric acid. When present in certain concentrations in aqueous solutions, proteinoids form small structures called microspheres, protobionts, or protocells. They bear much of the basic features provided by cell membranes. Evreinova et al. (1974) were able to produce stable protein-nucleic acid-carbohydrate coacervate drops stabilized by quinones which also absorb in the UVC (see table 1).

Proteinoid-based protocells enclosing DNA/RNA molecules could have been the first cellular life forms on Earth. The association of these amino acid proteinoid chains around a central RNA or DNA was again thermodynamically driven by the increase in the photon cross section offered by the amino acids taking advantage of the much superior dissipation characteristics of the RNA and DNA molecules, providing sub-pico second de-excitation of these photon excited aromatic amino acids to the ground state through internal conversion.

3. World of pigment-carrying protocells (early Archean eon, ~ 3.8 – 3.5 Ga)
Ancient (3.2 – 3.6 Ga) sedimentary rocks from the Pilbara Craton of Western Australia and the Barberton Greenstone Belt of South Africa and Swaziland have been found to contain prokaryotic microfossils (Noffke et al., 2013; Schopf, 2006). Evidence in these rocks suggests that 3.5 Ga old prokaryotic microorganisms flourished in the form of microbial mats and stromatolites. Stromatolites represent accretionary sedimentary structures from shallow water environments, produced by the activities of mat-building communities of mucilage-secreting microorganisms, mainly photoautotrophic prokaryotes (Schopf, 2006).

This antiquity of autotrophic (and possibly photosynthetic) activity is further corroborated by chemical markers like the ratio of $C^{13}/C^{12}$ in sedimentary organic carbon (kerogen), which indicates a continuous record of biological $CO_2$ fixation that goes back 3.5 - 3.8 Ga (Schidlowski, 1988; 2001). Among the earliest evidence for life on Earth is the biogenic graphite in 3.7 billion-year-old metasedimentary rocks discovered in Western Greenland (Ohtomo et al., 2014). Other evidence indicates that most of the biochemical pathways that drive modern prokaryotic carbon, sulfur and nitrogen cycles were in place by as early as 3.5 Ga, and by 2.7 Ga at the latest (Nisbet and Sleep, 2001; Grassineau et al., 2001).

The last universal common ancestor (LUCA) from which all organisms now living on Earth descend is estimated to have lived some 3.5 to 3.8 billion years ago (Glansdorff et al., 2008; Doolittle, 2000). It had properties currently shared by all three domains of life: cellular structure with water-based cytoplasm, enclosed by a lipid bilayer membrane; genetic code based on DNA or RNA; L-isomers of some of the 20 amino acids of proteins; ATP as an energy intermediate; common enzymes and cofactors, etc.

Since many more amino acids are chemically possible than the twenty found in modern protein molecules, and many other nucleotides are possible besides A, T, G, C, U, Theobald (2010) puts forward the possibility that LUCA was not alone, but only a member of an ancient, diverse microbial community. However, a strong thermodynamic reason, consistent with our proposition, for the existence of only the nominal nucleotides and their Watson-Crick pairing is that these have non-radiative de-excitation times orders of magnitude shorter than any other nucleotides and their possible pairings, even of those energetically more probable (Abo-Riziq et al., 2005; Woutersen and Cristalli, 2004).

From the optics of the thermodynamic dissipation theory for the origin of life (Michaelian, 2009; 2011), and the established facts given above, a picture of life on Earth prior 3.5 Ga can be inferred. From the multitude of different pigment complexes, the one that gave rise to the line of the LUCA evolved to be a prokaryote-like, lipid-membrane vesicle (protocell) where the 'old generation' of UV-dissipaters (DNA/RNA, some proteins) began to take on a new secondary role of information-carriers and chemical catalysts, becoming genetic material, ribosomes and enzymes, with a new function - to execute the synthesis, support, and proliferation of the 'new dissipating generation of pigments' more completely covering the evolving solar spectrum available at Earth's surface.

Cellular structure provided two new fundamental conditions: their increased photon cross section for dissipation, and protection of the progressively more intricate biochemical activity from the outside environment. Because increasing amounts of visible photons were making it to the surface as volcano produced sulfuric acid clouds similar to those on Venus today began to wane, it is logical to assume that the 'new generation' of dissipaters on the cell surface, as membrane-bound light-harvesting antenna molecules, were increasingly common towards absorption in the visible. Different types of organic pigments could have filled this role, but limiting our search to the line of the LUCA and keeping in mind the evidence of stromatolites and possible photosynthetic activity by this time, porphyrins/chlorins emerge as an interesting class of candidate pigments.

Many Urey-Miller-type experiments, using various sources of energy, have readily produced oligopyrroles and porphyrins abiotically, including using electrical discharges (Hodgson and Ponnamperuma, 1968; Simionescu et al., 1978), ultraviolet irradiation (Szutka, 1965; Hodgson and Baker, 1967; Meierhenrich et al. 2005), temperature of 70-100°C, compatible with a prebiotic milieu (Lindsey et al., 2009). Porphyrin-like substances have even been found in meteorites (Hodgson and Baker, 1964) and Precambrian rocks (Kolesnikov and Egorov, 1977). Porphyrins have a chemical affinity to DNA, their plane intercalates into DNA or they form stacked aggregates on its surface (Pasternack et al., 1991; Pasternack et al., 1993; Sigel and Sigel, 1996).

Siggel et al. (1996) have found that amphiphilic porphyrins aggregate in aqueous media and form fibers, ribbons, tubules and sheets. In lateral aggregates, the $S_x$ and $S_y$ components of the monomer Soret band centered around 400 nm are shifted to the blue and red respectively to yield exciton Soret bands at around 350 and 450 nm. There is substantial enhancement of internal conversion and thus quenching of fluorescence in these aggregates as compared to individual porphyrin molecules.

Chlorophyll (a chlorin derivative, related in structure to porphyrin) is also interesting in this respect. Isolated chlorophylls are susceptible to bleaching (the opening up of the tetrapyrrole ring) particularly at 350 nm where there is significant absorption. However, at high concentration they form aggregates that are particularly rapid quenchers from the UVC to the UVA and therefore stable against photoreactions and break-up (Zvezdanovic and Markovic, 2008). These same authors demonstrate that an accessory pigment environment has rather little effect on chlorophyll bleaching independent of the wavelength of the light, and therefore the assignation of accessory pigments as protectors of the photosynthetic system may be misguided. It is more likely that since the Archean, porphyrins, and later accessory pigments, formed part of an ever larger dissipative system, performing the most relevant thermodynamic dissipation associated to life, that of converting short wavelength photons into long wavelength photons.

Accordingly, primitive protocells could have acquired porphyrin molecules from UVC- and visible- dissipating aggregates in the environment, integrating them in the nucleic acid-protein complex as light-harvesting antenna pigments. The primitive cell then likely 'learned' very early in its evolution how to synthesize its own endogenic porphyrin, an assumption which can be backed by the transfer-RNA dependent aminolevulinic acid formation during chlorophyll biosynthesis in phototropic organisms, indicating porphyrin biosynthesis as ancient as that of proteins (Kumar et al., 1996; Grimm et al., 2006).

Further evidence for the antiquity of porphyrin-type pigments and their role in dissipation processes comes from the work of Mulkidjanian and Junge (1997). On the basis of evidence from sequence alignments between membrane-spanning segments of bacterial and plant proteins involved in photosynthesis (photosynthetic reaction centers and antenna complexes) they have postulated the existence of a large ancestral porphyrin-carrying protein, which they describe as a primordial UV-protector of the cell. UV-'protection' functioned in a way that the aromatic amino acid residues of the membrane protein clustered around the porphyrin/chlorin molecule were primary receivers of UV-quanta and transferred this excitation energy to the nearby pigment moiety which underwent rapid internal conversion to the lowest excited singlet state releasing the photon energy in thermal energy.

Chlorophyll, common in most modern phototropic organisms, plays essentially this dissipative role. It has two main excitation levels; the Soret band in the near UV and the low energy Qy band in the red part of the spectrum. The transition from the Soret into the Qy band with partial energy dissipation is very efficient, hence the enhancement of the Qy band and its shift to the red in the proposed evolutionary trajectory from protoporphyrin IX via protochlorophyll *c* and chlorophyll *c* to chlorophyll *a* (Olson and Pierson, 1987; Larkum, 1991).

Mulkidjanian and Junge (1997) further theorize how a purely dissipative photochemistry started in the context of UV-'protection' of the cell and later mutations, causing the loss of certain porphyrin-type pigments, led to acquisition of redox cofactors and a gradual transition from dissipative to productive photochemistry, i.e. from UV-'protection' to photosynthesis. In the context of the thermodynamic dissipation theory for the origin of life, this putative pigment-carrying membrane protein, instead of being an 'ancient UV-protector' is more appropriately designated as the 'new generation' of UV-dissipating pigments, continuing, from the nucleic and amino acids, the thermodynamic role of photon dissipation and catalysis of the water cycle. Since porphyrin/chlorin-carrying proteins are universal to all three domains of life (Georgopapadakou and Scott, 1977; Suo et al., 2007), either as chlorophyll-carrying photosystems, or heme-carrying cytochromes, it is quite plausible that this primordial porphyrin-protein complex was the main dissipating unit of the LUCA protocells.

Whether UV-based anoxygenic photosynthesis powered these primitive cells or they were dependent on chemoautotrophy, cannot be ascertained although previously mentioned evidence (chemical markers, stromatolites) imply that autotrophy and $CO_2$ fixation (possibly photoautotrophy) was well established by this time. The early emergence of autotrophy is consistent with the thermodynamic dissipation theory in the sense that it led to growth of organic mass, further pigment proliferation and hence still greater photon dissipation (Michaelian, 2009; 2011). Heterotrophs most likely developed somewhat later than autotrophs and were probably more prevalent in deeper waters feeding on the 'rain' of organic material from the surface. From a thermodynamic perspective they can be considered as 'gardeners' of the autotrophs, playing a crucial role in the constant recycling of the elements needed for pigment synthesis (phosphate, nitrogen, carbon etc.) and their expansion into new, initially inhospitable, areas.

4. World of pigment-carrying protocyanobacteria (Archean eon, ~ 3.5 – 2.5 Ga)
Despite the scarcity of Archean geologic units relative to those of the Proterozoic, the temporal distribution of stromatolites is more or less continuous from 3.5 to 2.5 Ga (Schopf et al., 2007). This distribution rather faithfully parallels the estimated temporal distribution of Archean sediments that have survived to the present, with most Archean stomatolites reported from rocks 3.0 to 2.5 Ga, where sedimentary rocks are relatively plentiful, and somewhat fewer reports from the older, 3.5 to 3.0 Ga interval. Biogenicity of these sediments has been backed by the discovery of microscopic fossils with microbe-like morphologies: small rod-shaped bodies, unornamented coccoids, or sinuous tubular or uniseriate filaments – morphologies typical of later, less questionable Proterozoic microfossils (Schopf, 2006). Tice and Lowe (2004; 2006) have provided geological evidence that the Buck Reef Chert (250 to 400 meter thick rock running along the South African coast) was produced by phototrophic microbial communities around 3.4 Ga ago. They also noted the absence of traces of life in the deeper (>200 m) water environments. Tice and Lowe defined the inhabitants of these primordial microbial communities as partially filamentous phototrophs, which, according to the carbon isotopic composition, used the Calvin–Benson–Bassham cycle to fix $CO_2$.

Combining this evidence with the results of Crowe et al. (2013) for the beginnings of atmosphere oxygenation would indicate that the advent of oxygenic photosynthesis was sometime prior 3 Ga, and was carried out by cyanobacteria-like organisms usually referred to in the scientific community as protocyanobacteria (Olson, 2006; Garcia-Pichel, 1998).

Cyanobacteria are believed to be one of the earliest branching groups of organisms on this planet (Altermann and Kazmierczak, 2003). They are the only known prokaryotes which carry out oxygenic photosynthesis, and there is little doubt that they played central role in the formation of atmospheric oxygen and ozone (Bekker et al., 2004).

Oxygenic photosynthesis is distinguished by the presence of two photosystems (photosystem I and photosystem II) of which photosystem II incorporates the water-splitting complex (four oxidized manganese atoms and a calcium atom, held in place by proteins), and the utilization of water as a reductant, which generates oxygen as a waste product (Kiang et al., 2007; Allen and Martin, 2007). In contrast, anoxygenic types of photosynthesis employ only one of the photosystems (photosystem I or photosystem II), but never both, lack the water-splitting complex and utilize other electron donors: hydrogen, hydrogen sulfide, ferrous ions, nitrite etc. (Kiang et al., 2007; Bryant and Frigaard, 2006).

Because of the far greater complexity of their photosynthetic machinery in comparison to other photosynthetic bacteria (purple bacteria, green sulfur bacteria, green non-sulfur bacteria, heliobacteria) cyanobacteria haven't been traditionally considered as a lineage in which photosynthesis could have emerged. But a recent study by Mulkidjanian et al. (2006, and Mulkidjanian and Galperin, 2013), comparing genome sequences of photosynthesis-related genes in phototropic members of diverse bacterial lineages, has come to the conclusion that photosynthesis originated in the cyanobacterial lineage and has spread to other bacterial phyla by way of lateral gene transfer. The study also proposes that this direct ancestor of the cyanobacteria – protocyanobacteria, was an anaerobic, mat-dwelling phototroph which probably used hydrogen as a reductant, and inherited its photosynthetic machinery from mutations of the earlier 'UV-protective' membrane protein. The invention of the water-splitting complex came later, but not before the lateral gene transfer to other bacteria which lack it, and, according to Mulkidjanian and Junge (1997), this complex was probably formed by the trapping of Mn-atoms in the cavities of earlier porphyrin-binding sites in photosystem II. Manganese ions of the type that occur in the water-splitting complex are readily oxidized by ultraviolet light of wavelengths less than 240 nm (Russel & Hall, 2006) and since there was no ozone to filter out the Sun's UV-rays before the origin of water splitting, these excited Mn atoms probably served as electron donors to protocyanobacteria which were introduced into an ancient, aquatic Mn-containing environment, and only later incorporated them into their photosystem II (Allen and Martin, 2007; Johnson et al., 2013). It is worth noting that a Mn-cluster containing electron vacancies might dissipate the energy of a single UV-quantum via formation of hydrogen peroxide from water (Mulkidjanian and Junge, 1997). The ability to produce dioxygen from water at the expense of four low-energy quanta of red light may have gradually developed at a later stage of protocyanobacterial evolution.

Dissipation of the continuingly intense UV-radiation during this geological period could have further been enhanced by the invention of an even greater variety of UV-absorbing compounds like the contemporary UV-A and UV-B absorbing mycosporine-like amino acids and scytonemin (Shick and Dunlap, 2002; Castenholz and Garcia-Pichel, 2002; Ferroni et al., 2010). These pigments, widespread in modern cyanobacteria, algae and other aquatic organisms and usually defined as UV-protectants or sunscreens, are most likely remnants of archean, protocyanobacterial, UV-dissipating pigments (Garcia-Pichel, 1998). Biosynthetically, mycosporine-like amino acids are derived from intermediates of the shikimate pathway for the synthesis of the aromatic amino acids (Shick and Dunlap, 2002; Portwich and Garcia-Pichel, 2003), while scytonemin is derived from tryptophan (Balskus et al., 2011), suggesting a later evolutionary appearance of these pigments compared to the aromatic amino acids.

Besides protocyanobacteria and other members of the domain Bacteria, the Archean seas were most likely also harboring the ancestors of modern domain Archaea (Wang et al., 2007; Woese and Gupta, 1981). If the ancestors of cyanobacteria were already present by 3 Ga, it is sensible to conclude that the line of the LUCA had already branched into the separate domains Bacteria and Archaea at a much earlier point in time, probably as early as 3.5 Ga, or even earlier. No known member of the Archaea domain carries out photosynthesis, although some members of the class Halobacteria are photoheterotrophs which utilize light energy absorbed by the chromoprotein bacteriorhodopsin to synthesize ATP, and their carbon source is organic (Bryant and Frigaard, 2006). Mulkidjanian and Junge (1997) hypothesize that the bacteriorhodopsin-based phototropism in Archaea might also have evolved from a 'UV-protecting' precursor function, because it has been shown (Kalisky et al., 1982) that a UV-quantum, channeled from the aromatic amino acid residues of the protein moiety, triggers the *all-tans* to 13-*cys* isomerization of the retinal moiety in bacteriorhodopsin, followed by its return into the initial isomeric state by slow thermal relaxation. This bacteriorhodopsin-based photon dissipation in Archaea probably evolved as a complementary to the chlorophyll-based in protocyanobacteria for dissipation of visible wavelengths left unabsorbed by chlorophyll (DasSarma, 2006, 2007).

5. World of pigment-carrying cyanobacteria and algal plastids (Proterozoic eon, ~ 2.5 Ga – 542 Ma)

Free oxygen is toxic to obligate anaerobic organisms and the rising concentrations during the early Proterozoic may have wiped out most of the Earth's anaerobic inhabitants at the time, causing one of the most significant extinction events in Earth's history known as the Oxygen Catastrophe (Sessions et al., 2009; Holland, 2006). Additionally the free oxygen reacted with the atmospheric methane, a greenhouse gas, reducing its concentration and thereby triggering the Huronian glaciation, possibly the most severe and longest snowball Earth episode (Kopp et al., 2005). Nevertheless, as soon as oxygen was available in the oceans, prokaryotes immediately discovered ways to utilize its power as an electron acceptor thus heralding the new age of aerobic respiration (Sessions et al., 2009; Raymond and Serge, 2006).

Biomarker evidence (the presence of long-chain 2-methylhopanes) suggests that during the early Proterozoic (Paleoproterozoic) a phytoplankton population was thriving in the ocean, consisting mainly, or entirely, of cyanobacteria (Summons et al., 1999; Falkowski et al., 2004; Canfield, 2005). Unlike other eubacterial phyla, cyanobacteria exhibit a substantially well-studied fossil record, with the earliest unequivocal cyanobacterial fossils dating back around 2.0 Ga which come from two localities, the Gunflint iron formation and the Belcher Subgroup, both in Canada (Sergeev et al., 2002; Golubic and Lee, 1999; Amard and Bertrand-Sarfati, 1997). Schirrmeister et al. (2013) suggest a concurrence of the onset of the Great Oxidation Event, the origin of cyanobacterial multicellular forms and an increased diversification rate of cyanobacteria.

The first advanced single-celled eukaryotes and multicellular life roughly coincide with the beginning of the accumulation of free oxygen in the atmosphere (El Albani et al., 2010; Falkowski and Isozaki, 2008; Baudouin-Cornu and Thomas, 2007). There is no consensus among biologists concerning the position of eukaryotes in the overall scheme of cell evolution (Brown, 2003). Autogenous models propose that originally a proto-eukaryotic cell containing a nucleus existed first, and later acquired mitochondria, which presumably came from the endosymbiosis of an aerobic proteobacterium engulfed by a proto-eukaryotic cell (Latorre et al., 2011; Doolittle, 2000), and it's generally assumed that all the eukaryotic lineages that did not acquire them went extinct (Martin and Mentel, 2010).

Chloroplasts most likely came about from another endosymbiotic event involving engulfed cyanobacteria into the primitive mitochondria-bearing eukaryotic cell; the available evidence indicates that their origin, too, was a singular event in evolution (Stoebe and Kowallik, 1999; McFadden, 2001), followed by a still uncertain number (between two and seven) of secondary endosymbiotic events in which a eukaryotic host engulfed a eukaryotic alga (Cavalier-Smith, 2000; Delwiche, 1999). Accordingly, the origins of mitochondria predate the origins of chloroplasts, but possibly not by much, such that mitochondria and eukaryotes might be about 2 billion years old, which is compatible with molecular clock estimates (Feng et al., 1997).

Chloroplasts are the organelles of plants and algae that harbor biochemical pathways for all pigments or pigment-precursors in the cell, like aromatic amino acids, heme, chlorophyll and isoprenoids (Keeling, 2004).
All primary chloroplasts belong to one of three chloroplast lineages - the glaucophyte chloroplast lineage, the rhodophyte (or red algal) lineage, and the chloroplastidan (or green algal) lineage (Ball et al., 2011). The second two are the largest, and the green chloroplast lineage is the one that contains land plants (Keeling, 2004). Glaucophyte chloroplasts contain photosynthetic pigments chlorophyll *a,* phycobilins, and phycobilisomes, small antenna-like structures organized on the outer face of thylakoid membranes that are also found in cyanobacteria (Keeling, 2004). Red algal chloroplasts contain chlorophyll *a*, phycobilins, and phycobilisomes, where the phycobiliprotein phycoerythrin is responsible for giving many red algae their distinctive red color. Green algal chloroplasts differ from glaucophyte and red algal chloroplasts in that they have lost their phycobilisomes, and contain chlorophyll *b* instead (Kim and Archibald, 2009). Phylogenetic analyses, however, show that the common ancestor of cyanobacteria and chloroplasts had both chlorophyll *b* and phycobilins (Tomitani et al., 1999).
The best supported molecular trees tend to favor the glaucophytes branching prior to the divergence of red and green algae from one another, which is specifically supported by two analyses of concatenated plastid- and nucleus-encoded genes (Martin et al., 1998; Moreira et al., 2000).

By about 1.5 Ga acritarchs become reasonably abundant fossil unicellular organisms that are almost certainly eukaryotes, and probably algae by virtue of an easily preserved cell wall (Javaux et al., 2001). By 1.2 Ga, very well-preserved multicellular red algae appear in the fossil record (Butterfield, 2000). This evidence further supports the theory that eukaryotes are at least 1.5 billion years old and implies that the diversification of the red algal lineage (which is not the most ancient lineage of algae) into multicellular forms occurred at least 1.2 billion years ago.

The end of the Proterozoic saw several, relatively short-lived, glaciation events, after which multicellular eukaryotic life boomed in the oceans (Ediacara biota) paving the way for the still greater boom during the Cambrian Explosion (Stanley, 2008).

Atmospheric changes during this eon led to the gradual dimming of UVC light available at the surface whose dissipation was being relegated from UVC-absorbing biological pigments to biologically-produced stratospheric ozone. The reduced intensity of UVC intercepted by the biosphere probably induced significant metabolic and pigment-related changes in the biological world. It is therefore conceivable that this was the period of diversification and proliferation of visible-absorbing pigments in cyanobacteria and their chloroplast counterparts in algae. Inside the photosynthetic apparatus, the work of primary photon receivers probably shifted from the aromatic amino acid residues of the light-harvesting antenna proteins (UVC-absorbers) to the (bacterio)chlorophyll pigments (visible and near-infrared absorbers). Contemporary carotenoid and phycobilin pigments, which function as accessory pigments to (bacterio)chlorophylls in photosynthesis, absorbing parts of the visible spectrum left unabsorbed by (bacterio)chlorophylls (Rowan, 1989; Nobel, 2009), probably became prevalent during this geological period.

Carotenoids (divided into carotenes and xanthophylls) are tetraterpenes (contain 8 isoprene units i.e. 40 carbon atoms) and because of the multiple conjugated double bonds in their polyene hydrocarbon chain, unlike UV-absorbing isoprene and related isoprenoids with less conjugated double bonds, their absorption maxima shifts to longer wavelengths in the blue-green (Rowan, 1989; Nobel, 2009). Phytoene, a 40-carbon intermediate in the biosynthesis of carotenoids, whose synthesis from two molecules of geranylgeranyl pyrophosphate (GGPP) marks the first committed step in the biosynthesis of carotenoids, has an UV-Vis absorption spectrum typical for a triply conjugated system with its main absorption maximum in the UVB range at 285 nm (Crounse et al., 1963). Phytofluene, the second product of carotenoid biosynthesis formed from phytoene in a desaturation reaction leading to the formation of five conjugated double bonds, has an absorption spectra in the UVA range, with maximal absorption at 348 nm (Crounse et al., 1963).

Phycobilins are linear tetrapyrroles, which could have easily evolve with the opening of the porphyrin macrocycle, as it happens during their biosynthetic pathway from heme (Beale, 1993; Brown et al., 1990). The light-harvesting antennas of cyanobacteria and the chloroplasts of red algae, glaucophytes and some cryptomonads consist of water-soluble protein-phycobillin complexes, known as phycobiliproteins (not present in green algae and higher plants) which efficiently absorb red, orange, yellow, and green wavelengths, not well absorbed by bacterio-chlorophylls (Rowan, 1989; Nobel, 2009). Ajlani and Vernotte (1998) have shown that phycobiliproteins are not essential for the survival of cyanobacteria by constructing a viable mutant of the cyanobacterium Synechocystis PCC 6803 which lacked major photosystem II phycobilin antenna. This phenomenon, seemingly inconsistent with Darwinian laws of evolution which state that organisms do not retain processes unbeneficial for their survival, finds consistency in the thermodynamic dissipation explanation of life if phycobiliproteins are seen more as dissipative structures instead of photosynthetic ones.

In addition to chlorophyll- and phycobilin-containing light-harvesting systems, cyanobacteria possess small chlorophyll-binding polypeptides, dubbed high-light-inducible polypeptides (HLIPs), which are not involved in photosynthesis and simply accumulate in the cell in response to intense light exposure and dissipate this energy to heat (Havaux et al., 2003; Montane and Kloppstech, 2000). This behavior has led biologists to ascribe a photo-protective role to this class of pigment-proteins, although we believe that their function might be more accurately described from the thermodynamic viewpoint as dissipaters of light and catalysts of the hydrologic cycle.

## 6. World of pigment-carrying cyanobacteria, algal and plant plastids (Phanerozoic eon, ~ 542 Ma - present)

The sudden emergence of many new species and phyla of animals during the Cambrian period (542 – 488 Ma) is known as the Cambrian explosion (Erwin and Valentine, 2013; Knoll and Carroll, 1999). Prokaryote lineages had probably colonized the land as early as 2.6 Ga, even before the origin of eukaryotes (Pisani et al., 2004), but for a long time the land remained barren of multicellular organisms. The oldest fossils of land fungi and plants date to 480-460 Ma, though molecular evidence suggests that fungi may have colonized the land as early as 1.0 Ga, and plants 700 Ma (Heckman et al., 2001). The spread of the organic pigments to land and their fomentation of a terrestrial water cycle extending ever more inland, provided the potential for another 30% of Earth's surface area to be cultivated for photon dissipation.

Land plants evolved from chlorophyte algae living in shallow fresh waters, perhaps as early as 510 million years ago (Raven and Edwards, 2001); some molecular estimates place their origin even earlier, as much as 630 million years ago (Clarke et al., 2011). Their chloroplasts contain the visible-absorbing pigments: chlorophyll *a* and *b*, different carotenes and xanthophylls, known as photosynthetic pigments (Ruban, 2012); and the visible- and UV-absorbing flavonoids, which are phenolic secondary plant metabolites that do not participate in photosynthesis and are traditionally given UV-protective and antioxidant roles (Agati et al., 2007; Andersen and Markham, 2010). Anthocyanins are flavonoid, water-soluble, vacuolar pigments, biosynthetically derived from phenylalanine, which occur in all tissues of higher plants, including leaves, stems, roots, flowers, and fruits. They absorb mainly in the green part of the spectrum, where chlorophylls and carotenoids do not absorb strongly, and may appear red, purple, or blue depending on the vacuolar pH (Gould et al., 2009). Betalains are a class of red and yellow pigments derived from tyrosine, found only in plants of the order Caryophyllales, where they replace anthocyanin pigments (Stafford, 1994).

Wang et al. (2007) have shown that transpiration rather than photosynthesis is the process maximized in plants. Transpiration is by far the largest water flux from Earth's continents, representing 80 to 90 % of terrestrial evapotranspiration (Jasechko et al., 2013). On the basis of analysis of a global data set of large lakes and rivers, Jasechko et al. conclude that transpiration recycles $62,000 \pm 8,000$ km$^3$ of water per year to the atmosphere, using half of all solar free energy available on the land surface. Being the energetically most consuming process in plants one can assume that the largest percent of photons absorbed by plant pigments are being dissipated (mainly through the process of nonphotochemical chlorophyll fluorescence quenching (NPQ) – Müller et al., 2001) and transferred to the breaking of hydrogen bonds between water molecules, which then evaporate through the stomata of leaves. In this fashion, terrestrial plants play the same role as phytoplankton (cyanobacteria, algae) over bodies of water (Michaelian, 2011; 2012a), and their evolution and colonization of land can be interpreted as the extension of entropy production through photon dissipation and the hydrologic cycle over land surfaces. As we have emphasized in this paper, the same thermodynamic function was performed by the earliest organic pigments floating on the surface of the primordial hadean ocean.

The timing of the first animals to leave the oceans is not precisely known: the oldest clear evidence is of arthropods on land around 450 Ma (Johnson et al., 1994), but there is also unconfirmed evidence that arthropods may have appeared on land as early as 530 Ma (MacNaughton et al., 2002).

The influence of insects, crustaceans and larger animals in spreading nutrients for the production of pigments, and thus the water cycle and photon dissipation, over barren land or mid ocean areas cannot be over emphasized (Michaelian, 2011). Doughty et al. (2013) have shown how the Pleistocene Amazonian megafauna extinctions dramatically decreased the lateral flux of nutrients (particularly of phosphorous) in all continents outside of Africa. Lavery et al. (2010) have determined that blue whales feces contribute enormously to the distribution of iron in the Southern Ocean for the production of cyanobacteria and thus photon dissipation and the sequestering of atmospheric carbon, the latter a very modern preoccupation.

## 4. The organic pigment tree of life correlates with the history of earth's surface spectrum

Summarizing the data presented in Section 2 for the abiotic and biotic events affecting Earth's atmosphere, a coarse grained history of the solar spectrum at Earth's surface can be derived. Before the origin of life at 3.85 Ga until about 3.2 Ga, the solar spectrum on Earth included a historically unique region of between approximately 240 nm and 290 nm with an integrated flux of $5 \pm 2$ $Wm^{-2}$ midday at the equator, with the spectral distribution in this region peaking at around 260 nm, bounded on the short wavelength side by $H_2S$ absorption and on the long wavelength side by formaldehyde and acetaldehyde (both photochemical reaction products of UV on hydrogen sulfide). From 290 nm to about 320 nm, little light would have been available on the surface due to absorption of the aldehydes. Beyond 320 nm until about 700 nm, where the water absorption becomes important (a warmer Earth would have supported a much more humid atmosphere), the spectrum at Earth's surface in the Archean was probably very similar to the present day spectrum but of about 30% lower luminosity due to a faint young sun, providing an integrated energy flux of approximately 327 $Wm^{-2}$ integrated over 320 to 700 nm midday at the equator. The biological world in this era consisted of the dominance of UVC-dissipating pigments in protocells, among them LUCA, and included the split of the line of the LUCA into Bacteria and Archaea and the early branching of the protocyanobacterial line within Bacteria.

From 3.2 Ga to 2.7 Ga, a thick upper atmosphere organic haze probably existed due to the rise of the methanogens, bringing the $CH_4/CO_2$ ratio close to one, which is the condition for haze production (Low and Tice, 2004). The wavelength dependence of the resulting albedo would depend on the size of the agglomerates formed in the atmosphere, probably increasing that of longer wavelengths by a relatively greater amount (as on Titan today; McGrath et al., 1998), but would have generally increased the albedo at all wavelengths with an ensuing cooling of Earth's surface locally to glacier temperatures. The biosphere saw the branching of oxygenic cyanobacteria within the protocyanobacterial line and beginning of gradual $O_2$ accumulation with the continuing dominance of UV-dissipating pigments.

From about 2.7 to 2.45 Ga the organic haze produced by far UV light on $CH_4$ in the upper atmosphere began to decline as erosion of the continents and tectonic recycling of $CO_2$ reduced the $CH_4/CO_2$ ratio, thereby increasing the intensity of the spectrum at Earth's surface between 320 and 700 nm while at the same time ozone, resulting from oxygenic photosynthesis, was beginning to reduce the intensity of the UV spectrum between 240 to 290 nm. Dissipation in this wavelength range was now becoming ever more relegated to life derived ozone in the upper atmosphere. The biosphere was seeing the diversification of cyanobacteria and the appearance of multicellular forms of cyanobacteria, increasingly dissipating in the visible.

After about 2.45-2.2 Ga ozone from oxygenic photosynthesis would have reduced light within the 240-290 nm wavelength range to less than perhaps a few percent of its value at the beginnings of life ($5\pm2$ Wm$^{-2}$ at 3.85 Ga). The biosphere was experiencing the endosymbiosis of mitochondria and chloroplasts and the emergence of eukaryotes. Visible-dissipating pigments probably became dominant in this era.

The above summary of the solar spectrum evolution at Earth's surface is depicted in figures 2 and 3 where at 3.85 Ga we have included the absorption of $H_2S$ and aldehydes, Rayleigh scattering for atmospheric pressures at an estimated 1.5 times present day values, and water at roughly twice its average concentration of present day values of 5 mm of precipitable water (Gates, 1980).

The calculated spectrum at 2.9 Ga includes absorption of $H_2S$ and aldehydes, Rayleigh scattering for atmospheric pressures at about 1.2 times present day values, and water at roughly 1.5 its average present concentration, plus wavelength independent scattering due to organic haze increasing the albedo to 0.6.

At 2.2 Ga we include absorption due to oxygen at 0.15 of present day mass concentration of 21%, giving an absolute value of 3.2 %, and ozone also at 0.15 of present day concentrations, Rayleigh scattering for atmospheric pressure at 1.2 present day values, water at 1.2 of its average present day concentration, and scattering due to organic haze at about 0.10 of its late Archean concentration.

For the spectrum calculation at present day, 0.0 Ga, we include absorption of oxygen at 21%, ozone absorption and Rayleigh scattering at present atmospheric pressures giving an integrated energy flow of 865 Wm$^{-2}$ midday at the equator, consistent with measured values (Gates, 1980).

All calculations assumes a present day blackbody solar photosphere at an effective temperature of 5840 K with an H$^-$ stellar atmosphere absorption line centered on 500 nm, and includes the increase in the solar radius and increase in surface temperature over time as considered in section 2.1 above.

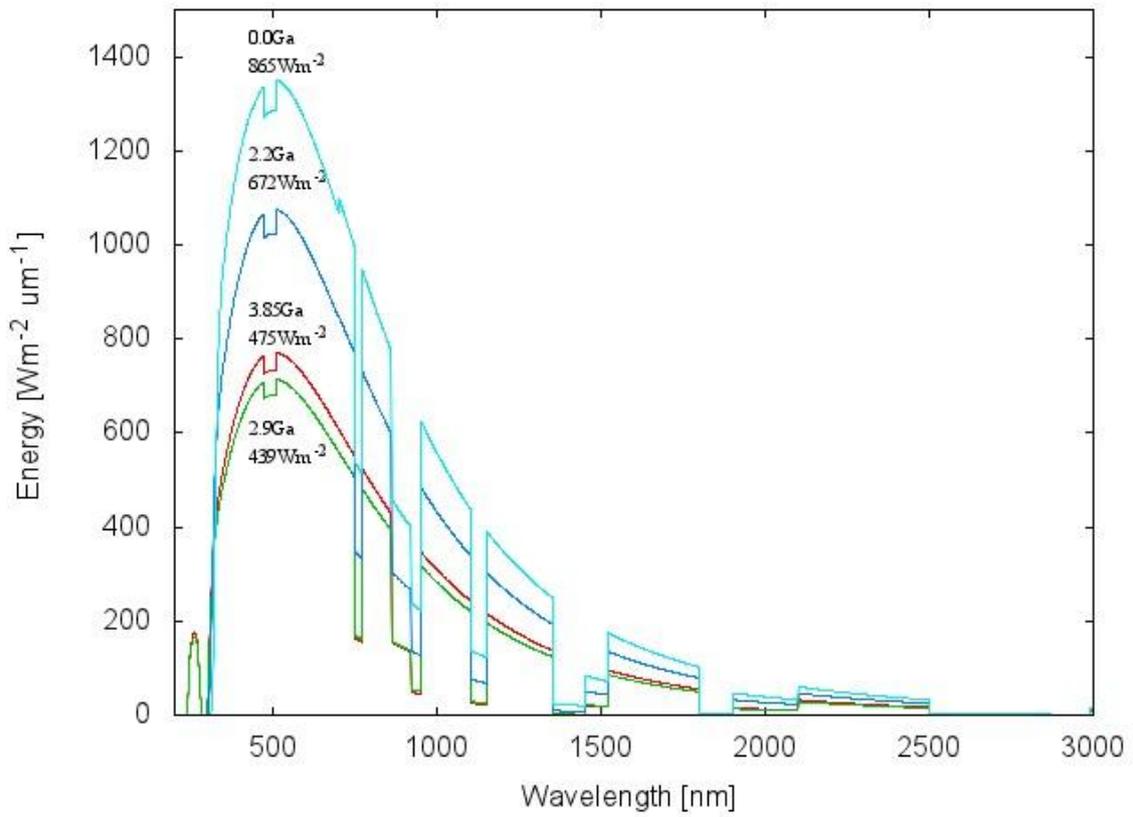

Fig. 2 Calculated solar spectra at Earth's surface at particular dates since present using the best available data for the atmospheric gasses, their densities, and surface temperature. The total integrated (over wavelength) energy flows midday at the equator are given below the dates.

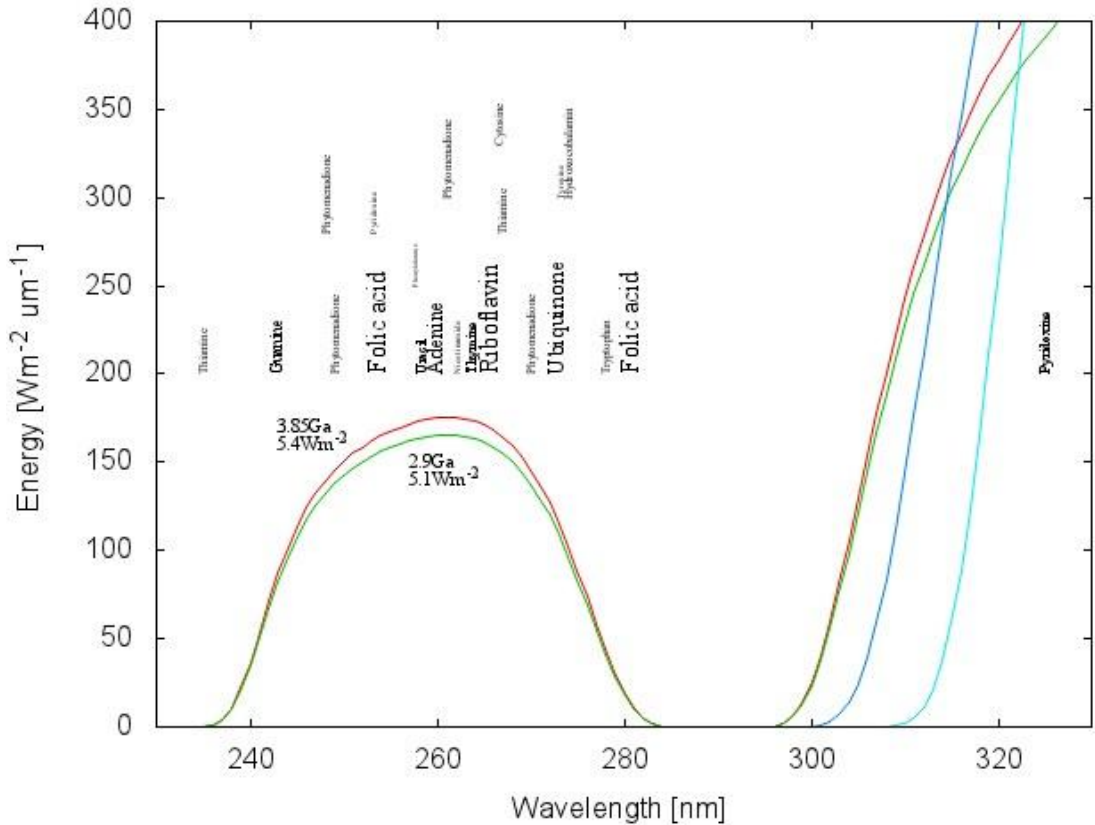

Fig. 3 Locations of maximum absorption of primary pigments in wavelength coincide with the predicted solar spectrum at Earth's surface in the UVC at the time of the origin of life, 3.85 Ga. The font size of the letter is roughly proportional to the absorption cross section of the indicated pigment.

The calculations presented in figures 2 and 3 have internal consistency with; an increase of about 27% in the integrated energy flow arriving at the top of Earth's atmosphere since the beginnings of life; a total of 865 $Wm^{-2}$ arriving at Earth's surface today, and with 9% of the energy flow in the ultraviolet below 400 nm. The calculation gives 5.4 $Wm^{-2}$ within the range 230-290 nm arriving at Earth's surface and 4.3 $Wm^{-2}$ within the reduced wavelength range 240-270 nm at the beginning of life (3.85 Ga), roughly consistent with Sagan's (1973) calculation of 3.3 $Wm^{-2}$.

Figure 3 plots the first organic pigments (primary molecules of life) which are common to all three domains of life (see table I) at their absorption maxima. All pigments dissipate strongly in the 230-290 nm range, just where a window existed in Earth's Archean atmosphere. The wavelength range of absorption for each pigment is in fact much wider than can be depicted in the figure, and the asymmetry around peak absorption in most cases is also substantial. For example, guanine actually absorbs strongly over the whole 240-280 nm range but has peak absorption at 243 nm.

The correlation between the absorption spectra of the primary molecules of life and the solar spectrum at Earth's surface during the Archean is supportive of the thermodynamic dissipation theory of the origin of life (Michaelian, 2009, 2011) which suggests that life arose and evolved to dissipate the solar photon flux and would have started out oriented towards dissipation of the shorter wavelengths, not merely because these are the most photochemically active wavelengths, but because this is the region which gives greatest entropy production per unit photon, and per unit energy, upon conversion to a black-body spectrum at the temperature of Earth's surface.

## 5. Conclusions

Given the current understanding of the evolution of solar like stars and a likely, but less certain, evolution of Earth's atmosphere, we have determined the probable history of the solar spectrum at Earth's surface. The most notable feature of this analysis is the existence, since before the beginnings of life at 3.85 Ga until at least 2.7 Ga, of a UV light component between 230 and 290 nm of $5\pm2$ Wm$^{-2}$. There is some direct empirical evidence for this UV atmospheric window in the form of $^{33}$S isotope inclusions in diamonds dated at 2.9 Ga that could only have been formed through specific atmospheric photochemical reactions with ultraviolet light in this wavelength range (Farquhar et al., 2002). This light could have provided the free energy for a number of photochemical reactions leading to complex organic pigments which are fundamental molecules of life including RNA, DNA, amino acids, enzymatic cofactors, quinones and porphyrins.

These fundamental bio-molecules, common to all three domains of life, absorb and dissipate strongly within the UVC range of 230 to 290 nm, especially when intercalated or coupled externally to RNA and DNA which acts as a quencher molecule. A non-linear, non-equilibrium thermodynamic directive suggests a spontaneous proliferation of these dissipating molecules to concentrations much greater than their expected near equilibrium concentration over the entire surface of the Earth exposed to UVC radiation. This is simply explained by the fact that the production of these pigments can be seen as an autocatalytic photochemical reaction process established to dissipate the solar photon potential (Michaelian, 2013).

The proliferation and evolution of these pigment molecules over time, which succinctly characterizes all biolotic and coupled biotic-abiotic evolution (Michaelian, 2012b), can best be described through five basic tendencies; to increase the ratio of their effective photon absorption cross sections to their physical size, to decrease their electronic excited state life times, to quench radiative de-excitation channels (e.g. fluorescence), to cover ever more completely the solar spectrum, and finally to disperse into an ever greater surface area of Earth. The appearance of mobile animals and protozoa can be seen to foment this thermodynamic directive which has the overall objective of increasing the entropy production of Earth in its solar environment.

These primary molecules today have lost their direct dissipating utility and are produced only through complex metabolic reactions employing special enzymes and special free energy sources, however, they are still completely relevant in their supportive role that they play in fomenting the proliferation and distribution of the new pigments over Earth's surface.

This is important evidence in support of the thermodynamic dissipation theory of the origin of life (Michaelian, 2009, 2011) which states that life arose and proliferated to carry out the thermodynamic function of dissipating the entropically most important part of the solar spectrum (the shortest wavelength photons) available at Earth's surface and that this irreversible process began to evolve and couple with other irreversible abiotic processes, such as the water cycle, to become more efficient, to cover ever more of the solar spectrum, and ever more of Earth's surface. This dissipation continues today as the most important thermodynamic function performed by living organisms as part of a greater dissipating system known as "the biosphere" (Michaelian, 2012b).

Finally, our suggestion provides new avenues for exploration of Earth's ancient atmosphere based on pigment history analysis and sheds new light on the origin of photosynthesis. For example, the lack of primordial pigments absorbing between 290 and 320 nm strongly suggest the presence of aldehydes in the archean atmosphere as postulated by Sagan (1973). The demonstrated abiotic production utilizing UV light of porphyrins, their UVC absorption and dissipation characteristics when stacked as agglomerates and intercalated or coupled externally to RNA or DNA, suggests that photosynthesis arose as an addition to their UV light-harvesting function and autocatalytic proliferation when this photochemically active light at Earth's surface began to wane as life produced ozone took on the dissipative role at these wavelengths in the upper atmosphere. A greater in depth analysis remains to be performed on the relation between RNA/DNA and porphyrins and that between dissipation and photosynthesis and we are presently engaged in experiments to explore these ideas.

**Acknowledgements**
K.M. is grateful for the financial support provided by DGAPA-UNAM project number IN-103113.

# G

## .J

## K

# T